\documentclass[12pt,a4paper]{article}
\usepackage{graphicx}
\usepackage{hyperref}
\usepackage{a4wide}
\usepackage{amsmath}
\usepackage[numbers,square,sort&compress]{natbib}
\allowdisplaybreaks[1]

\begin{document}

\begin{titlepage}
\begin{flushright}
LU TP 16-44\\
August 2016
\end{flushright}
\vfill
\begin{center}
{\Large\bf Pion light-by-light contributions to the muon $g-2$}
\vfill
{\bf Johan Bijnens and Johan Relefors}\\[0.3cm]
{Department of Astronomy and Theoretical Physics, Lund University,\\
S\"olvegatan 14A, SE 223-62 Lund, Sweden}\\[2mm]

\end{center}
\vfill
\begin{abstract}
This paper contains some new results on the hadronic light-by-light
contribution (HLbL) to the muon $g-2$. 
The first part argues that we can expect large effects from disconnected
diagrams in present and future calculations by lattice QCD of HLbL.
The argument is based on the dominance of pseudo-scalar meson
exchange.

In the second part, we
revisit the pion loop HLbL contribution to the muon anomalous
magnetic moment.
We study it in the framework of some models studied earlier, pure pion loop,
full VMD and hidden local symmetry for inclusion of vector mesons.
In addition we study possible ways to include the axial-vector meson.
The main part of the work is a detailed study of how the
different momentum regions contribute.
We derive a short distance constraint on the $\gamma^*\gamma^*\to\pi\pi$
amplitude and use this as a constraint on the models used for the pion loop.
As a byproduct we present the general result for integration using the
Gegenbauer polynomial method.
\end{abstract}
\vfill
\end{titlepage}

\section{Introduction}

The muon anomalous magnetic moment is one of the most precise measured
quantities in high energy physics. The muon anomaly measures the
deviation of the magnetic moment away from the prediction of a Dirac point
particle
\begin{equation}
a_\mu \equiv \frac{g_\mu-2}{2}\,.
\end{equation}
where $g_\mu$ is the gyromagnetic ratio $\vec M = g_\mu (e/2m_\mu)\vec S$.
The most recent experiment at BNL \cite{gm2exp1,gm2exp2,gm2exp3,PDG}
obtains the value
\begin{equation}
a_\mu = 11\ 659\ 208.9(5.4)(3.3)~10^{-10}\,,
\end{equation}
an impressive precision of 0.54~ppm (or 0.3~ppb on $g_\mu$). The new experiment at Fermilab
aims to improve this precision to 0.14~ppm \cite{gm2expFNAL} and
there is a discussion whether a precision of 0.01~ppm is feasible
\cite{gm2expJPARC}. In order to fully exploit the reach of these experiments
an equivalent precision needs to be reached by the theory.
The theoretical prediction consist of three main parts,
the pure QED contribution, the electroweak contribution and
the hadronic contribution. 
\begin{equation}
a_\mu = a_\mu^\mathrm{QED}+a_\mu^\mathrm{EW}+a_\mu^\mathrm{had}\,.
\end{equation}
An introductory review of the theory is
\cite{Knechtlectures} and more comprehensive review are \cite{Miller:2007kk,JNreview}.
Recent results can be found in the proceedings of the
conferences \cite{talk3,Proceedings:2016bng}.

\begin{figure}
\centerline{
\includegraphics[width=12cm]{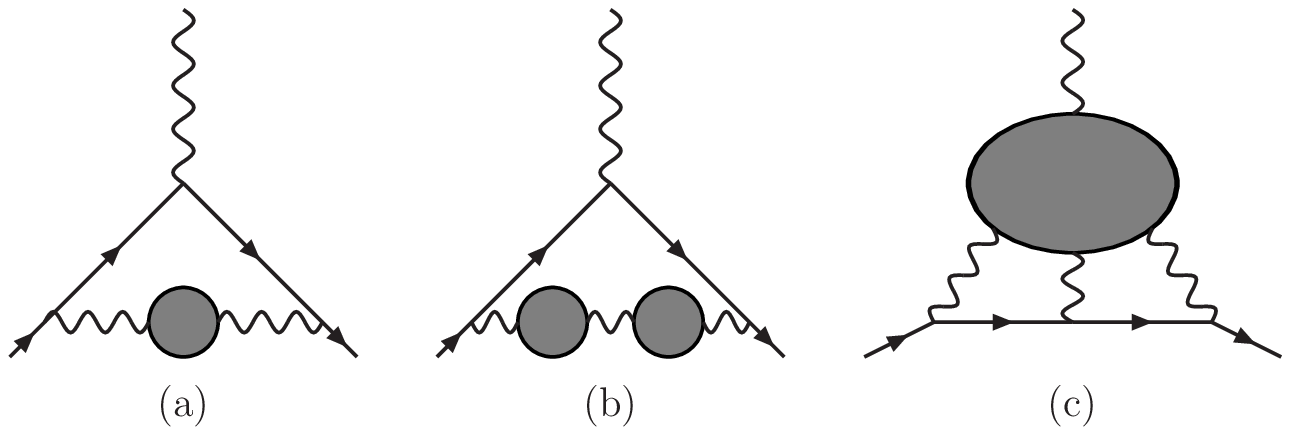}
}
\caption{\label{fighadronic}
The thee main hadronic contributions to the muon anomalous
magnetic moment. (a)The lowest order hadronic vacuum polarization. (b)
An example of a higher order hadronic vacuum polarization contribution.
(c) The light-by-light scattering contribution. In all three cases
the shaded regions represent the hadronic part.}
\end{figure}

The hadronic part has two different contributions, those due to hadronic
vacuum polarization, both at lowest and higher orders, and the light-by-light
scattering contributions. 
\begin{equation}
a_\mu^\mathrm{had} = a_\mu^\mathrm{LO\mbox{-}HVP}+a_\mu^\mathrm{HO\mbox{-}HVP}
+a_\mu^\mathrm{HLbL}\,.
\end{equation}
These are depicted symbolically in
Fig.~\ref{fighadronic}.

The hadronic vacuum polarization contributions can be related to the
experimentally measured cross-section $e^+e^-\to$~hadrons. Here the accuracy can
thus in principle be improved as needed for the experimental
measurements of $a_\mu$.

The more difficult light-by-light contribution has no such simple relation
to experimentally measurable quantities. A first comprehensive calculation
appeared in \cite{KNO}. One of the main problems there was the possibility
of double counting when comparing quark-loop, hadron-loop and hadron exchange
contributions. A significant step forward was done
when it was realized \cite{deRafael} that the different contributions
start entering at a different order in the expansion in the number of colours
$N_c$ and in the chiral power counting, order in momentum $p$.
This splitting was then used by two groups to estimate the
light-by-light contribution \cite{HKS1,HKS2,HK}(HKS)
and \cite{BPP1,BPP2,BPP3}(BPP).
After correcting a sign mistake made by both groups for different reasons
and discovered by \cite{KN} the
results are
\begin{equation}
a_\mu^\mathrm{HLbL} =  8.96(1.54)~10^{-10}~~(HKS),
\qquad  8.3(3.2)~10^{-10}~~(BPP)\,.
\end{equation}
A new developments since then have been the inclusion of
short distance constraints on the full correction \cite{MV}(MV)
which indicated a larger contribution
\begin{equation}
a_\mu^{HLbL} = 13.6(2.5)~10^{-10}~~(MV)\,.
\end{equation}
Comparisons in detail of the various contributions in these three
main estimates can be found in \cite{BP} and \cite{PRV}.
An indication of a possibly larger quark-loop
contribution are the recent Schwinger-Dyson estimates of that contribution
\cite{SDE1,SDE2,SDE3,SDE4}. First results of using dispersion relations to get
an alternative handle on HLbL have also appeared
\cite{Colangelo:2014dfa,Colangelo:2014pva,Colangelo:2015ama,Pauk:2014rfa}.
Lattice QCD has now started to contribute to HLbL as well, see e.g.
\cite{Blum:2015gfa,Green:2015sra} and references therein.

In this paper we add a number of new results to the HLbL discussion.
First, in Sect.~\ref{disconnected} we present an argument why in the
lattice calculations the disconnected contribution is expected to be large
and of opposite sign to the connected contribution. This has been
confirmed by the first lattice calculation \cite{Jin}.
The second part is extending the Gegenbauer polynomial method to
do the integration over the photon momenta \cite{KN,JNreview}
to the most general
hadronic four-point function. This is the subject of Sect.~\ref{method}.
The third and largest part 
is about the
charged pion and kaon loop. These have been estimated rather differently in the
the three main evaluations
\begin{equation}
\label{piloop1}
a_\mu^{\pi loop} =
-0.45(0.81)~10^{-10}~(HKS),\quad
-1.9(1.3)~10^{-10}~(BPP),\quad
0.0(1.0)~10^{-10}~(MV).
\end{equation}
The numerical result is always dominated by the charged pion-loop, the
charged kaon loop is about 5\% of the numbers quoted in (\ref{piloop1}).
The errors in all cases were mainly the model dependence. The main goal of
this part is to show how these differences arise in the calculation
and include a number of additional models. Given the uncertainties we will
concentrate on the pion-loop only.

There are several improvements in this paper over the previous work
on the pion loop.
First, we use the Gegenbauer polynomial method of \cite{JNreview,KN}
to do two more of the integrals analytically compared to the earlier work.
Second, we study more models by including the vector mesons in a number
of different ways and study the possible inclusion of axial-vector mesons.
That the latter might introduce some uncertainty has been emphasized in
\cite{Engel:2012xb,Engel:2013kda}. We present as well a new short-distance
constraint that models have to satisfy for the underlying
$\gamma\gamma\pi\pi$ vertex.

Our main tool for understanding the different results is to study the dependence
on the virtualities of the three internal photons in Fig.~\ref{fighadronic}(c).
The use of this as a method to understand contributions was started
in \cite{BP} for the main pion exchange. One aspect that will become clear is
that one must be very careful in simply adding more terms in a hadronic model.
In general, these models are non-renormalizable and there is thus no
guarantee that there is a prediction for the muon anomaly in general. In fact,
we have not found a clean way to do it for the axial vector meson as discussed
in Sect.~\ref{piloop}. However, using that the results should have a decent 
agreement with ChPT at low energies and the high-energy constraint and only integrating up to a reasonable hadronic scale we obtain the result
\begin{equation}
a_\mu^{HLbL~\pi loop} = -(2.0\pm0.5)\cdot 10^{-10}\,.
\end{equation}
This is discussed in Sect.~\ref{piloop}.

A short summary is given in Sect.~\ref{Conclusions}.
Some of the results here have been presented earlier in \cite{talk3,talk1,talk2}
and \cite{Mehranthesis}.

\section{Large disconnected contributions}
\label{disconnected}

\begin{figure}
\centerline{\includegraphics[width=8cm]{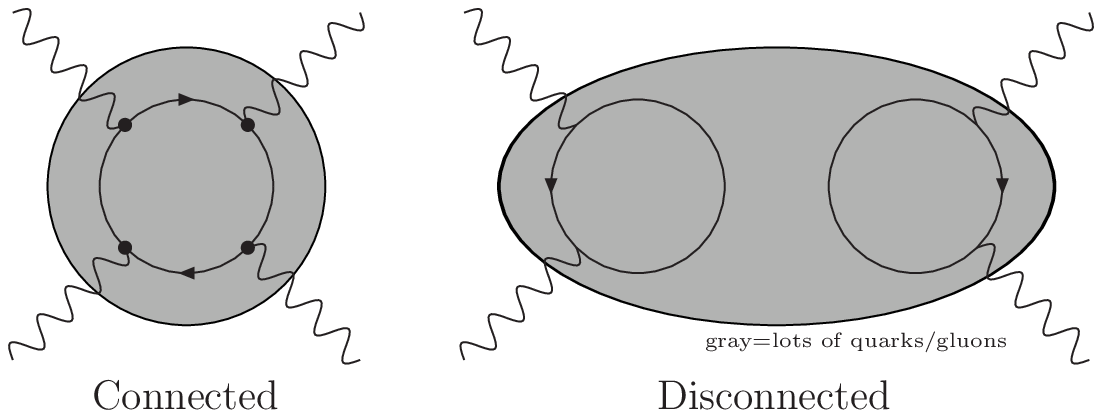}}
\caption{
\label{figdisconnected}
The connected contribution where all photons couple to a single quark-loop
and an example of a disconnected diagram where the photons couple to different
quark-loops.}
\end{figure}

Lattice calculations of HLbL are starting to give useful results. One question
here is how to calculate the full contribution including both connected
and disconnected contributions. The latter is more difficult to calculate,
see e.g. \cite{Hayakawa:2015ntr}, and many calculations so far have only
presented results for the connected contribution. In this section we
present an argument why the disconnected contribution is expected to be large
and of opposite sign to the connected contribution. The connected contribution
is the one where the four photons present in Fig.~\ref{fighadronic}(c) all
connect to the same quark line, the disconnected contribution where they
connect to different quark lines. This is depicted schematically in
Fig.~\ref{figdisconnected}.
The argument below is presented for the case of two-flavours and has been
presented shortly in \cite{talk2}.

A large part of the HLbL contribution comes from pseudo-scalar meson exchange.
For that part of the contribution we can give some arguments on the relative
size of the disconnected and connected contribution.
An example of a limit where the connected contribution is the only one
is the large $N_c$ limit. One important consequence of this limit is that
the anomalous breaking of the $U(1)_A$ symmetry disappears and the flavour
singlet pseudo-scalar meson becomes light as well. This also
applies to exchanges of other multiplets, but there the mass differences between
the singlet and non-singlet states are much smaller.

Let us first look at the quark-loop case with two flavours.
The connected diagram
has four photon couplings, thus each quark flavour gives a contribution
proportional to its charge to the power four. The connected contribution has
thus a factor of $q_u^4+q_d^4=(2/3)^4+(-1/3)^4=17/81$.
For the disconnected contribution we have instead charge factors of the
form $(q_u^2+q_d^2)$ for each quark-loop, so the final result has a factor
of $(q_u^2+q_d^2)^2=25/81$. However, this does not give any indication of the
relative size since the contributions are very different.

\begin{figure}
\centerline{
\begin{minipage}{4cm}
\includegraphics[width=\textwidth]{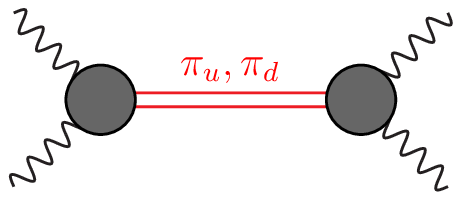}
\centerline{(a)}
\end{minipage}
\hspace{1cm}
\begin{minipage}{4cm}
\includegraphics[width=\textwidth]{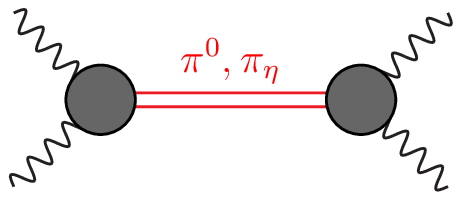}
\centerline{(b)}
\end{minipage}
}
\caption{\label{figpi0} The meson-exchange picture. 
(a) With $\pi_u$ and $\pi_d$ exchange. (b) With $\pi^0$ and $\pi_\eta$ exchange.
}
\end{figure}

In the large $N_c$ limit the mesons are the flavour eigenstates.
We then have two light neutral pseudo-scalars,
one with flavour content $\bar u u$, $\pi_u$ and one with $\bar d d$, $\pi_d$.
In the meson exchange picture, shown in Fig.~\ref{figpi0}(a) the coupling
of $\pi_u$ to two photons is proportional to $q_u^2$, thus $\pi_u$ exchange
has factor of $q_u^4$. The same argument goes for the $\pi_u$ exchange
and we obtain a factor of $q_d^4$. The total contribution is thus
proportional to $q_u^4+q_d^4=17/81$ in agreement with the quark-loop
argument for the same contribution.

We can also work with the isospin eigenstates instead. These are the $\pi^0$
with flavour content $(\bar u u-\bar d d)/\sqrt{2}$ and the flavour singlet
$\pi_\eta$ with flavour content $(\bar u u+\bar d d)/\sqrt{2}$.
In the large $N_c$ limit we should obtain the same result as with $\pi_u$ and
$\pi_d$. The $\pi^0$ coupling to 2 photons is proportional to
$\delta_{\pi^0}=(q_u^2-q_d^2)/\sqrt{2}=3/(9\sqrt{2})$. The $\pi_\eta$ coupling
to two photons is $\delta_{\pi_\eta}=(q_u^2+q_d^2)/\sqrt{2}=5/(9\sqrt{2})$.
The exchange of $\pi^0$ and $\pi_\eta$ leads to a contribution proportional to
$\delta^2_{\pi^0}+\delta^2_{\pi_\eta}=17/81$ in agreement with the argument from
the quark-loop or $\pi_u,\pi_d$ exchange.

What happens now if we turn on the disconnected contribution or remove the
large $N_c$ limit. The 
physical eigenstates are now $\pi_\eta$ and $\pi^0$
and they no longer have the same mass. In effect, from the breaking of the
$U(1)_A$ the singlet has gotten a large mass and its contribution becomes much
smaller. In the limit of being able to neglect $\pi_\eta$-exchange completely
the sum of connected and disconnected contributions is reproduced by $\pi^0$
exchange alone which is proportional to $\delta_{\pi^0}^2=(9/2)/81$.
So in this limit we expect the total contribution
is $\delta_{\pi^0}^2$ times a factor $A$. From the discussion in the
previous paragraph follows that the connected part is
$\delta_{\pi^0}^2+\delta_{\pi_\eta}^2$ times the same factor $A$.
The disconnected part must thus cancel the $\delta_\pi^2$ part of the
connected contribution and must be
$-\delta_{\pi_\eta}^2$ times again the factor $A$.
We thus expect a large and negative disconnected
contribution with a ratio of disconnected to connected
of $-25/34$.

There are really three flavours $u,d,s$ to be considered but the argument
generalizes straightforward to that case with
case $\delta_{\pi^0}=3/(9\sqrt{2})$, $\delta_\eta = 3/(9\sqrt{6})$
and $\delta_{\eta^\prime}=6/(9\sqrt{3})$.
In the equal mass case the ratio of disconnected to connected is for three
flavours
$-\delta{\eta^\prime}^2/(\delta_{\pi^0}^2+\delta_\eta^2+\delta_{\eta^\prime}^2)
= -2/3$.

The above argument is valid in the equal mass limit, assuming the
singlet does not contribute after $U(1)_A$ breaking is taken into account
and only for the pseudo-scalar meson-exchange. There are corrections following
from all of these. For most other contributions the disconnected effect
is expected to be smaller. The ratio of disconnected to connected of $-2/3$
is thus an overestimate but given that $\pi^0$ exchange is the largest
contribution we expect large and negative disconnected contributions.

Note that the above argument was in fact already used in the pseudo-scalar
exchange estimate of \cite{BPP1,BPP2,BPP3}, the comparison
of the large $N_c$ estimate and $\pi^0,\eta,\eta^\prime$ exchange is in
Table 2 and the separate contributions in Table 3 of \cite{BPP2}, up to the
earlier mentioned overall sign.

Lattice QCD has been working hard on including disconnected contributions
\cite{Hayakawa:2015ntr}. Using the same method of \cite{Blum:2015gfa}
at physical pion mass preliminary results were shown at Lattice 2016
\cite{Jin} of $11.60(96)$ for the connected and $-6.25(80)$ for the
disconnected in units of $10^{-10}$. This is in good agreement with the
arguments given above.

\section{The Gegenbauer polynomial method}
\label{method}

\begin{figure}
\centerline{
\includegraphics[width=6cm]{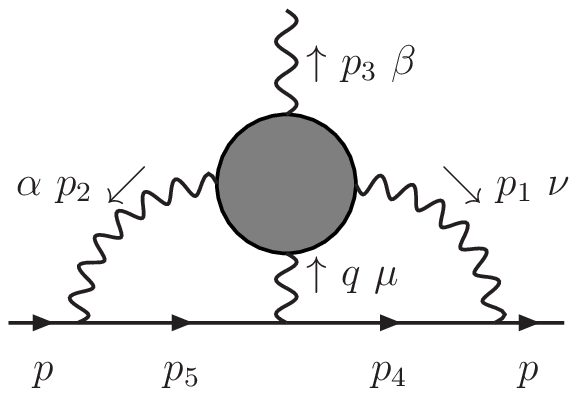}
}
\caption{\label{figrouting}The momentum routing for the muon line
and through the hadronic four-point function as used in
(\ref{defMlb}).}
\end{figure}

The hadronic light-by-light contribution to the muon anomalous magnetic
moment is given by \cite{ABDK}
\begin{equation}
\label{projector}
a_\mu^{LbL} =\frac{-1}{48 m_\mu}\mathrm{tr}
\left[\left(p\hskip-0.98ex/+m_\mu\right)
M^{\lambda\beta}(0)\left(p\hskip-0.98ex/+m_\mu\right)
\left[\gamma_\lambda,\gamma_\beta\right]
\right]\,,
\end{equation}
with
\begin{equation}
\label{defMlb}
M^{\lambda\beta}(p_3) = e^6
\int\frac{d^4 p_1}{(2\pi)^4}\frac{d^4 p_2}{(2\pi)^4}
\frac{\gamma_\nu\left(p_4\hskip-1.9ex/\hskip0.4ex+m\right)\gamma_\mu
\left(p_5\hskip-1.9ex/\hskip0.4ex+m\right)\gamma_\alpha}
{q^2 p_1^2 p_2^2 \left(p_4^2-m^2\right)\left(p_5^2-m^2\right)}
\left[\frac{\partial}{\partial p_{3\lambda}}
\Pi^{\mu\nu\alpha\beta}\left(p_1,p_2,p_3\right)\right]\,.
\end{equation}
Here $m$ is the muon mass, $p$ is the muon momentum, $q=p_1+p_2+p_3$,
$p_4=p-p_1$ and $p_5=p+p_2$.
The momentum routing in the diagram is shown in Fig.~\ref{figrouting}.
Note that because of charge conjugation the integration in
(\ref{defMlb}) is symmetric under the interchange of $p_1$ and $p_2$.
The symmetry under the full interchange of $-q,p_1,p_2$ is only
explicitly present if the other permutations of the photons on the muon
line are also added and then averaged. In this manuscript we stick to
using only the permutation shown. The integral gives still the
full contribution because the different permutations are included in
the hadronic four-point function $\Pi^{\mu\nu\alpha\beta}(p_1,p_2,p_3)$.

The hadronic four-point function is
\begin{equation}
\label{defPI}
\Pi^{\mu\nu\alpha\beta}(p_1,p_2,p_3)
= i^3\int d^4x d^4y d^4z
e^{i(p_1\cdot x+p_2\cdot y+p_3 \cdot z)} 
\langle0|T\left(V^\mu(0)V^\nu(x)V^\alpha(y)V^\beta(z)\right)|0\rangle\,.
\end{equation}
The current is $V_\mu = \sum_q Q_q \bar q \gamma_\mu q$ with $q$ denoting the
quarks and $Q_q$ the quark charge in units of $|e|$.
The four-point function has a rather complicated structure
and we discuss this in more detail Sect.~\ref{fourpoint}.

The partial derivative in (\ref{defMlb}) was introduced by \cite{ABDK}
to make each photon leg permutation of the fermion-loop finite which allows
to do the numerical calculation at $p_3=0$. It used
$p_{3\beta} \Pi^{\mu\nu\alpha\beta} = 0$ to obtain via $\partial/\partial p_{3\lambda}$
\begin{equation}
0 = \Pi^{\mu\nu\alpha\lambda}+p_{3\beta}\frac{\partial}{\partial p_{3\lambda}}
\Pi^{\mu\nu\alpha\beta}\,.
\end{equation}

The integral in (\ref{defMlb}) contains 8 degrees of freedom.
After projecting on the muon magnetic moment with (\ref{projector}) it can only
depend on $p_1^2, p_2^2, p_1\cdot p_2, p\cdot p_1, p\cdot p_2$. The earlier work
in \cite{HKS1,HKS2,HK,BPP1,BPP2,BPP3} relied on doing all these integrals
numerically and in \cite{BPP1,BPP2,BPP3} this was done after an additional
rotation to Euclidean space. For the pion exchange contribution a method was
developed to reduce the number of integrals from 5 to 2 using the method of
Gegenbauer polynomials \cite{KN}. The assumptions made there about the behaviour
of the hadronic four-point function are not valid for the parts we study in
this paper. However, in \cite{JNreview} for the pion and scalar exchange
contributions the same method
has been used to explicitly perform the integrals over the
$p\cdot p_1$ and $p\cdot p_2$ degrees of freedom.
The same method can be used
to perform the integral over these two degrees of freedom also in the case
for the most general four-point function. This leads to an expression of
about 260 terms expressed in the combinations \cite{BPP2} 
of the four point function that contribute to the muon $g-2$.
We have checked that our calculation reproduces for the pion exchange the
results quoted in \cite{JNreview}.

\subsection{The general four-point function}
\label{fourpoint}

The four-point functions defined in (\ref{defPI}) contains
138 different Lorentz-structures \cite{BPP2}\footnote{Note that this
is the most general case also valid in other dimensions.
For four dimensions there are some additional constraints leading to
only 136 independent components \cite{SDE4}. This is not relevant
for the work presented here.}
\begin{align}
\Pi^{\mu\nu\alpha\beta}(p_1,p_2,p_3) &\equiv  \Pi^{1}(p_1,p_2,p_3)
 g^{\mu\nu} g^{\alpha\beta} +
\Pi^{2}(p_1,p_2,p_3) g^{\mu\alpha} g^{\nu\beta}
\nonumber\\
&+\Pi^{3} (p_1,p_2,p_3)
g^{\mu\beta} g^{\nu\alpha} \nonumber \\
&+\Pi^{1jk}(p_1,p_2,p_3)
 g^{\mu\nu} p_j^\alpha p_k^\beta +
\Pi^{2jk}(p_1,p_2,p_3)
 g^{\mu\alpha} p_j^\nu p_k^\beta \nonumber \\
&+ \Pi^{3jk}(p_1,p_2,p_3)
 g^{\mu\beta} p_j^\nu p_k^\alpha +
\Pi^{4jk}(p_1,p_2,p_3)
 g^{\nu\alpha} p_j^\mu p_k^\beta \nonumber \\
&+ \Pi^{5jk}(p_1,p_2,p_3)
 g^{\nu\beta} p_j^\mu p_k^\alpha +
\Pi^{6jk}(p_1,p_2,p_3)
 g^{\alpha\beta} p_j^\mu p_k^\nu \nonumber \\
&+ \Pi^{ijkm}(p_1,p_2,p_3)
 p_i^\mu p_j^\nu p_k^\beta p_m^\alpha \, ,
\end{align}
where $i,j,k,m =$ 1, 2 or 3 and repeated indices are summed.
The functions are scalar functions of all possible invariant products
$p_i\cdot p_j$.

The four point function satisfies the
Ward-Takahashi identities
\begin{equation}
\label{Ward}
q_\mu\Pi^{\mu\nu\alpha\beta}=
p_{1\nu}\Pi^{\mu\nu\alpha\beta}=
p_{2\alpha}\Pi^{\mu\nu\alpha\beta}=
p_{3\beta}\Pi^{\mu\nu\alpha\beta}=0\,.
\end{equation}
These identities allow to show that there are 43 independent functions
in general. Of course, since the four-point function is symmetric under
the interchange of the external legs many of these are related by permutations.

In practice it is easier not to do this reduction, but only the partial step
up to reducing them to the 64 functions $\Pi^{ijkm}$. This can be done such that
the powers of $p_3$ appearing explicitly never decrease.
Not all of these
contribute to $a_\mu$, in fact at most 32 combinations can contribute
\cite{BPP2}. These are the $\Pi^{3jkm}, \Pi^{i3km}, \Pi^{ij3m}$
and the $\Pi^{Dijk}$, all with $i,j,k=1,2$.
The  $\Pi^{Dijk}$ come from derivatives of the $\Pi^{ijkm}$ w.r.t. $p_{3\lambda}$
at $p_3=0$
\begin{align}
\frac{\partial}{\partial p_{3\lambda}}\Pi^{ijkm}
&= p_1^\lambda \Pi^{1ijkm}+p_2^\lambda\Pi^{2ijkm}\nonumber\\
\Pi^{Dijk} &= \Pi^{1ijk2}-\Pi^{2ijk1}\,.
\end{align}

\subsection{The Gegenbauer method}

The simplification introduced in \cite{KN} was that the Gegenbauer polynomial
method can be used to average over all directions of the muon momentum.
After this averaging is done there is only dependence on the
invariant quantities $p_1^2, p_2^2$ and $p_1\cdot p_2$ left.
The method is fully explained in \cite{JNreview}. One can apply it
to the full four-point function or to the one where one has reduced the number
of components by using the Ward identities to the 64 $\Pi^{ijkl}$.

So we first take (\ref{projector}) and (\ref{defMlb}) and rotate everything
to Euclidean momenta $P_1$, $P_2$ and $P$ with $Q=P_1+P_2$, $P_4=P-P_1$
and $P_5=P+P_2$.
We see that the muon momentum $P$ shows up in denominators
with $p_4^2-m^2 = -(P_4^2+m^2)$ and $p_5^2-m^2=-(P_5^2+m^2)$ only.
After taking the Dirac trace only scalar products of momenta are
present in the numerator.
Removing the products $P\cdot P_1$ and $P\cdot P_2$ by completing them
to the full $P_4^2+m^2$ and $P_5^2+m^2$, the angular averaging over muon momenta
can be performed using \cite{JNreview}
\begin{eqnarray}
\left\langle \frac{1}{(P_4^2+m^2)(P_5^2+m^2)} \right\rangle_\mu
&=& \delta X\,,
\nonumber\\
\left\langle \frac{P\cdot P_1}{P_5^2+m^2} \right\rangle_\mu
&=& \frac{1}{8}\delta P_1\cdot P_2 r_2^2\,,
\nonumber\\
\left\langle \frac{P\cdot P_2}{P_4^2+m^2} \right\rangle_\mu
&=& \frac{1}{8}\delta P_1\cdot P_2 r_2^2\,,
\nonumber\\
\left\langle \frac{1}{P_4^2+m^2} \right\rangle_\mu
&=& \frac{1}{2} \delta r_1\,,
\nonumber\\
\left\langle \frac{1}{P_5^2+m^2} \right\rangle_\mu
&=& \frac{1}{2} \delta r_2\,.
\end{eqnarray}
Here we used the notation
\begin{eqnarray}
\label{defX}
\delta &=&\frac{1}{m^2}\,,
\nonumber\\
r_i &=& 1-\sqrt{1+\frac{4m^2}{P_i^2}}
\nonumber\\
X &=& \frac{1}{P_1 P_2\sin\theta}\mathrm{atan}\left(\frac{z \sin\theta}{1-z\cos\theta}
\right)
\nonumber\\
\cos\theta &=& \frac{P_1\cdot P_2}{P_1 P_2}
\nonumber\\
z&=&\frac{P_1 P_2}{4 m^2}r_1 r_2\,.
\end{eqnarray}

The final contribution to the muon anomaly is given by
\begin{equation}
\label{defintegral}
a_\mu = \frac{\alpha^3}{2\pi^2}\int
P_1^2 dP_1^2 P_2^2 dP_2^2 \sin\theta d\cos\theta A_\Pi(P_1,P_2,\cos\theta)\,.
\end{equation}
The quantity $A_\Pi$ is given by
\begin{align}
\label{resultgegenbauer}
       \Pi^{1131}& ( 
          - 1/6 \, \rho_3^2 r_2^2 \delta
          - 2/3 \, \rho_1 \rho_3 r_2 \delta
          + 8/3 \, \rho_1 \rho_3 X
          -     \, \rho_1^2 r_1 \delta
          - 4/3 \, \rho_1^2 \rho_3 X \delta
          - 2   \, \rho_1^2 \rho_2 X \delta
          )\nonumber\\
     + \Pi^{1132}&  (
          + 2/3 \rho_3
          + 1/3 \rho_2 \rho_3 r_2 \delta
          - 1/6 \rho_2 \rho_3 r_2^2 \delta
          - 2/3 \rho_1 \rho_3 r_1 \delta
          - 1/6 \rho_1 \rho_3 r_1^2 \delta
          - 2/3 \rho_1 \rho_2 r_2 \delta
\nonumber\\&
          + 1/3 \rho_1 \rho_2 r_1 \delta
          + 8/3 \rho_1 \rho_2 X
          - 4/3 \rho_1 \rho_2 \rho_3 X \delta
          + 2/3 \rho_1 \rho_2^2 X \delta
          - 4/3 \rho_1^2 \rho_2 X \delta
          )\nonumber\\
       + \Pi^{1231}   &(
          - 2/3 \rho_3^2 r_2 \delta
          - 1/6 \rho_2 \rho_3 r_2^2 \delta
          - 2/3 \rho_1 \rho_3 r_1 \delta
          - 4/3 \rho_1 \rho_3^2 X \delta
          + 1/3 \rho_1 \rho_2 r_2 \delta
\nonumber\\&
          + 8/3 \rho_1 \rho_2 X
          - 4/3 \rho_1 \rho_2 \rho_3 X \delta
          + 2/3 \rho_1^2 \rho_2 X \delta
          )\nonumber\\
       + \Pi^{1232}   &(
          - 2/3 \rho_3^2 r_1 \delta
          - 2/3 \rho_2
          - 2/3 \rho_2 \rho_3 r_2 \delta
          + 8/3 \rho_2 \rho_3 X
          - 4/3 \rho_2 \rho_3^2 X \delta
          - 1/3 \rho_2^2 r_2 \delta
\nonumber\\&
          - 1/3 \rho_1 \rho_2 r_1 \delta
          - 4/3 \rho_1 \rho_2 \rho_3 X \delta
          - 2/3 \rho_1 \rho_2^2 X \delta
          )\nonumber\\
       + \Pi^{1311}   &(
          + 1/3 \rho_1 \rho_3 r_2 \delta
          + 1/3 \rho_1^2 r_1 \delta
          + 2/3 \rho_1^2 \rho_3 X \delta
          + 2/3 \rho_1^2 \rho_2 X \delta
          )\nonumber\\
       + \Pi^{1312}   &(
          - 2/3 \rho_3^2 r_2 \delta
          + 4/3 \rho_3^2 X
          - 1/12 \rho_2 \rho_3 r_2^2 \delta
          - 4/3 \rho_1 \rho_3 r_1 \delta
          - 1/12 \rho_1 \rho_3 r_1^2 \delta
\nonumber\\&
          - 4/3 \rho_1 \rho_3^2 X \delta
          + 1/2 \rho_1 \rho_2 r_2 \delta
          + 1/6 \rho_1 \rho_2 r_1 \delta
          + 4/3 \rho_1 \rho_2 X
          - 8/3 \rho_1 \rho_2 \rho_3 X \delta
\nonumber\\&
          + 1/3 \rho_1 \rho_2^2 X \delta
          + \rho_1^2 \rho_2 X \delta
          )\nonumber\\
       + \Pi^{1322}   &(
          - 2/3 \rho_2
          - 2/3 \rho_2 \rho_3 r_2 \delta
          + 8/3 \rho_2 \rho_3 X
          - 1/3 \rho_2^2 r_2 \delta
          - 2 \rho_1 \rho_2 r_1 \delta
\nonumber\\&
          - 4/3 \rho_1 \rho_2 \rho_3 X \delta
          - 4 \rho_1 \rho_2^2 X \delta
          )\nonumber\\
       + \Pi^{2131}   &(
          - 2/3 \rho_1
          - 2/3 \rho_1 \rho_3 r_1 \delta
          + 8/3 \rho_1 \rho_3 X
          - 2 \rho_1 \rho_2 r_2 \delta
          - 4/3 \rho_1 \rho_2 \rho_3 X \delta
          - 1/3 \rho_1^2 r_1 \delta
\nonumber\\&
          - 4 \rho_1^2 \rho_2 X \delta
          )\nonumber\\
       + \Pi^{2231}   &(
          - 2/3 \rho_3^2 r_1 \delta
          + 4/3 \rho_3^2 X
          - 4/3 \rho_2 \rho_3 r_2 \delta
          - 1/12 \rho_2 \rho_3 r_2^2 \delta
          - 4/3 \rho_2 \rho_3^2 X \delta
          - 1/12 \rho_1 \rho_3 r_1^2 \delta
\nonumber\\&
          + 1/6 \rho_1 \rho_2 r_2 \delta
          + 1/2 \rho_1 \rho_2 r_1 \delta
          + 4/3 \rho_1 \rho_2 X
          - 8/3 \rho_1 \rho_2 \rho_3 X \delta
          + \rho_1 \rho_2^2 X \delta
          + 1/3 \rho_1^2 \rho_2 X \delta
          )\nonumber\\
       + \Pi^{2232}   &(
          + 1/3 \rho_2 \rho_3 r_1 \delta
          + 1/3 \rho_2^2 r_2 \delta
          + 2/3 \rho_2^2 \rho_3 X \delta
          + 2/3 \rho_1 \rho_2^2 X \delta
          )\nonumber\\
       + \Pi^{2311}   &(
          - 2/3 \rho_3^2 r_2 \delta
          - 2/3 \rho_1
          - 2/3 \rho_1 \rho_3 r_1 \delta
          + 8/3 \rho_1 \rho_3 X
          - 4/3 \rho_1 \rho_3^2 X \delta
          - 1/3 \rho_1 \rho_2 r_2 \delta
\nonumber\\&
          - 4/3 \rho_1 \rho_2 \rho_3 X \delta
          - 1/3 \rho_1^2 r_1 \delta
          - 2/3 \rho_1^2 \rho_2 X \delta
          )\nonumber\\
       + \Pi^{2312}   &(
          - 2/3 \rho_3^2 r_1 \delta
          - 2/3 \rho_2 \rho_3 r_2 \delta
          - 4/3 \rho_2 \rho_3^2 X \delta
          - 1/6 \rho_1 \rho_3 r_1^2 \delta
          + 1/3 \rho_1 \rho_2 r_1 \delta
          + 8/3 \rho_1 \rho_2 X
\nonumber\\&
          - 4/3 \rho_1 \rho_2 \rho_3 X \delta
          + 2/3 \rho_1 \rho_2^2 X \delta
          )\nonumber\\
       + \Pi^{2321}   &(
          + 2/3 \rho_3
          - 2/3 \rho_2 \rho_3 r_2 \delta
          - 1/6 \rho_2 \rho_3 r_2^2 \delta
          + 1/3 \rho_1 \rho_3 r_1 \delta
          - 1/6 \rho_1 \rho_3 r_1^2 \delta
          + 1/3 \rho_1 \rho_2 r_2 \delta
\nonumber\\&
          - 2/3 \rho_1 \rho_2 r_1 \delta
          + 8/3 \rho_1 \rho_2 X
          - 4/3 \rho_1 \rho_2 \rho_3 X \delta
          - 4/3 \rho_1 \rho_2^2 X \delta
          + 2/3 \rho_1^2 \rho_2 X \delta
          )\nonumber\\
       + \Pi^{2322}   &(
          - 1/6 \rho_3^2 r_1^2 \delta
          - 2/3 \rho_2 \rho_3 r_1 \delta
          + 8/3 \rho_2 \rho_3 X
          - \rho_2^2 r_2 \delta
          - 4/3 \rho_2^2 \rho_3 X \delta
          - 2 \rho_1 \rho_2^2 X \delta
          )\nonumber\\
       + \Pi^{3111}   &(
          + 1/6 \rho_3^2 r_2^2 \delta
          - 2/3 \rho_1
          - 4/3 \rho_1 \rho_3 r_2 \delta
          + 1/2 \rho_1 \rho_3 r_2^2 \delta
          - 1/3 \rho_1 \rho_2 r_2 \delta
          - \rho_1^2 r_2 \delta
\nonumber\\&
          - 1/3 \rho_1^2 r_1 \delta
          - 8/3 \rho_1^2 \rho_3 X \delta
          - 2/3 \rho_1^2 \rho_2 X \delta
          - 2 \rho_1^3 X \delta
          )\nonumber\\
       + \Pi^{3112}   &(
          + 4/3 \rho_3
          + 2/3 \rho_2 \rho_3 r_2 \delta
          + 1/6 \rho_2 \rho_3 r_2^2 \delta
          + 2/3 \rho_1
          + 2/3 \rho_1 \rho_3 r_1 \delta
          - 1/3 \rho_1 \rho_3 r_1^2 \delta
\nonumber\\&
          - 8/3 \rho_1 \rho_3 X
          + 2/3 \rho_1 \rho_2 r_1 \delta
          - 8/3 \rho_1 \rho_2 X
          + 4/3 \rho_1 \rho_2 \rho_3 X \delta
          + 4/3 \rho_1 \rho_2^2 X \delta
          + 1/3 \rho_1^2 r_1 \delta
          )\nonumber\\
       + \Pi^{3121}   &(
          + 2 \rho_1
          + \rho_1^2 r_1 \delta
          )\nonumber\\
       + \Pi^{3122}   &(
          + 2 \rho_2
          + \rho_2^2 r_2 \delta
          )\nonumber\\
       + \Pi^{3211}   &(
          + 4/3 \rho_3
          - 8/3 \rho_3^2 X
          + 2/3 \rho_2 \rho_3 r_2 \delta
          + 2/3 \rho_1
          + 2/3 \rho_1 \rho_3 r_1 \delta
          - 1/6 \rho_1 \rho_3 r_1^2 \delta
\nonumber\\&
          - 8/3 \rho_1 \rho_3 X
          + 1/3 \rho_1 \rho_2 r_2 \delta
          + 1/3 \rho_1 \rho_2 r_1 \delta
          + 4/3 \rho_1 \rho_2 \rho_3 X \delta
          + 2/3 \rho_1 \rho_2^2 X \delta
\nonumber\\&
          + 1/3 \rho_1^2 r_1 \delta
          + 2/3 \rho_1^2 \rho_2 X \delta
          )\nonumber\\
       + \Pi^{3212}   &(
          + 4/3 \rho_3
          - 8/3 \rho_3^2 X
          + 2/3 \rho_2
          + 2/3 \rho_2 \rho_3 r_2 \delta
          - 1/6 \rho_2 \rho_3 r_2^2 \delta
          - 8/3 \rho_2 \rho_3 X
\nonumber\\&
          + 1/3 \rho_2^2 r_2 \delta
          + 2/3 \rho_1 \rho_3 r_1 \delta
          + 1/3 \rho_1 \rho_2 r_2 \delta
          + 1/3 \rho_1 \rho_2 r_1 \delta
          + 4/3 \rho_1 \rho_2 \rho_3 X \delta
\nonumber\\&
          + 2/3 \rho_1 \rho_2^2 X \delta
          + 2/3 \rho_1^2 \rho_2 X \delta
          )\nonumber\\
       + \Pi^{3221}   &(
          + 4/3 \rho_3
          + 2/3 \rho_2
          + 2/3 \rho_2 \rho_3 r_2 \delta
          - 1/3 \rho_2 \rho_3 r_2^2 \delta
          - 8/3 \rho_2 \rho_3 X
          + 1/3 \rho_2^2 r_2 \delta
          + 2/3 \rho_1 \rho_3 r_1 \delta
\nonumber\\&
          + 1/6 \rho_1 \rho_3 r_1^2 \delta
          + 2/3 \rho_1 \rho_2 r_2 \delta
          - 8/3 \rho_1 \rho_2 X
          + 4/3 \rho_1 \rho_2 \rho_3 X \delta
          + 4/3 \rho_1^2 \rho_2 X \delta
          )\nonumber\\
       + \Pi^{3222}   &(
          + 1/6 \rho_3^2 r_1^2 \delta
          - 2/3 \rho_2
          - 4/3 \rho_2 \rho_3 r_1 \delta
          + 1/2 \rho_2 \rho_3 r_1^2 \delta
          - 1/3 \rho_2^2 r_2 \delta
          - \rho_2^2 r_1 \delta
          - 8/3 \rho_2^2 \rho_3 X \delta
\nonumber\\&
          - 2 \rho_2^3 X \delta
          - 1/3 \rho_1 \rho_2 r_1 \delta
          - 2/3 \rho_1 \rho_2^2 X \delta
          )\nonumber\\
       + \Pi^{D111}   &(
          - 1/3 \rho_1 \rho_3
          + 2/3 \rho_1 \rho_3^2 X
          - 1/6 \rho_1 \rho_2 \rho_3 r_2 \delta
          + 1/24 \rho_1 \rho_2 \rho_3 r_2^2 \delta
          - 1/6 \rho_1^2 \rho_3 r_1 \delta
\nonumber\\&
          + 1/24 \rho_1^2 \rho_3 r_1^2 \delta
          - 1/12 \rho_1^2 \rho_2 r_2 \delta
          - 1/12 \rho_1^2 \rho_2 r_1 \delta
          - 2/3 \rho_1^2 \rho_2 X
          - 1/3 \rho_1^2 \rho_2 \rho_3 X \delta
\nonumber\\&
          - 1/6 \rho_1^2 \rho_2^2 X \delta
          - 1/6 \rho_1^3 \rho_2 X \delta
          )\nonumber\\
       + \Pi^{D121}   &(
          + 1/3 \rho_3^2
          - 2/3 \rho_3^3 X
          + 1/6 \rho_2 \rho_3^2 r_2 \delta
          - 1/24 \rho_2 \rho_3^2 r_2^2 \delta
          + 1/6 \rho_1 \rho_3^2 r_1 \delta
          - 1/24 \rho_1 \rho_3^2 r_1^2 \delta
\nonumber\\&
          + 1/12 \rho_1 \rho_2 \rho_3 r_2 \delta
          + 1/12 \rho_1 \rho_2 \rho_3 r_1 \delta
          + 2/3 \rho_1 \rho_2 \rho_3 X
          + 1/3 \rho_1 \rho_2 \rho_3^2 X \delta
          + 1/6 \rho_1 \rho_2^2 \rho_3 X \delta
\nonumber\\&
          + 1/6 \rho_1^2 \rho_2 \rho_3 X \delta
          )\nonumber\\
       + \Pi^{D122}   &(
          + 2/3 \rho_2 \rho_3
          - 4/3 \rho_2 \rho_3^2 X
          + 1/3 \rho_2^2 \rho_3 r_2 \delta
          - 1/12 \rho_2^2 \rho_3 r_2^2 \delta
          + 1/3 \rho_1 \rho_2 \rho_3 r_1 \delta
\nonumber\\&
          - 1/12 \rho_1 \rho_2 \rho_3 r_1^2 \delta
          + 1/6 \rho_1 \rho_2^2 r_2 \delta
          + 1/6 \rho_1 \rho_2^2 r_1 \delta
          + 4/3 \rho_1 \rho_2^2 X
          + 2/3 \rho_1 \rho_2^2 \rho_3 X \delta
\nonumber\\&
          + 1/3 \rho_1 \rho_2^3 X \delta
          + 1/3 \rho_1^2 \rho_2^2 X \delta
          )\nonumber\\
       + \Pi^{D211}   &(
          - 2/3 \rho_1 \rho_3
          + 4/3 \rho_1 \rho_3^2 X
          - 1/3 \rho_1 \rho_2 \rho_3 r_2 \delta
\nonumber\\&
          + 1/12 \rho_1 \rho_2 \rho_3 r_2^2 \delta
          - 1/3 \rho_1^2 \rho_3 r_1 \delta
          + 1/12 \rho_1^2 \rho_3 r_1^2 \delta
          - 1/6 \rho_1^2 \rho_2 r_2 \delta
          - 1/6 \rho_1^2 \rho_2 r_1 \delta
\nonumber\\&
          - 4/3 \rho_1^2 \rho_2 X
          - 2/3 \rho_1^2 \rho_2 \rho_3 X \delta
          - 1/3 \rho_1^2 \rho_2^2 X \delta
          - 1/3 \rho_1^3 \rho_2 X \delta
          )\nonumber\\
       + \Pi^{D221}   &(
          - 1/3 \,\rho_3^2
          + 2/3 \,\rho_3^3 X
          - 1/6 \,\rho_2 \rho_3^2 r_2 \delta
          + 1/24\, \rho_2 \rho_3^2 r_2^2 \delta
          - 1/6 \,\rho_1 \rho_3^2 r_1 \delta
          + 1/24\, \rho_1 \rho_3^2 r_1^2 \delta
\nonumber\\&
          - 1/12\, \rho_1 \rho_2 \rho_3 r_2 \delta
          - 1/12\, \rho_1 \rho_2 \rho_3 r_1 \delta
          - 2/3 \,\rho_1 \rho_2 \rho_3 X
          - 1/3 \,\rho_1 \rho_2 \rho_3^2 X \delta
\nonumber\\&
          - 1/6 \,\rho_1 \rho_2^2 \rho_3 X \delta
          - 1/6 \,\rho_1^2 \rho_2 \rho_3 X \delta
          )\nonumber\\
       + \Pi^{D222}   &(
          + 1/3 \, \rho_2 \rho_3
          - 2/3 \, \rho_2 \rho_3^2 X
          + 1/6 \, \rho_2^2 \rho_3 r_2 \delta
          - 1/24\,  \rho_2^2 \rho_3 r_2^2 \delta
          + 1/6 \, \rho_1 \rho_2 \rho_3 r_1 \delta
\nonumber\\&
          - 1/24\,  \rho_1 \rho_2 \rho_3 r_1^2 \delta
          + 1/12\,  \rho_1 \rho_2^2 r_2 \delta
          + 1/12\,  \rho_1 \rho_2^2 r_1 \delta
          + 2/3 \, \rho_1 \rho_2^2 X
          + 1/3 \, \rho_1 \rho_2^2 \rho_3 X \delta
\nonumber\\&
          + 1/6 \, \rho_1 \rho_2^3 X \delta
          + 1/6 \, \rho_1^2 \rho_2^2 X \delta
          )\,.
\end{align}
Here we used the abbreviations 
$\rho_1 = P_1^2$, $\rho_2 = P_2^2$ and $\rho_3 = P_1\cdot P_2$.
in addition to those defined above.

A more general formula without using the Ward identities can also be derived.
Quoting this one would be too long. In practice for many models, the
method without using Ward identities leads to shorter but equivalent
results.
We have used both options for the bare pion loop, the full VMD (Vector Meson
Dominance) model
and the hidden local symmetry (HLS) model and only the latter method for the
antisymmetric field model for the vector and axial vector mesons.

\section{The pion-loop contribution to HLbL}
\label{piloop}

\begin{figure}
\centerline{\includegraphics[width=12cm]{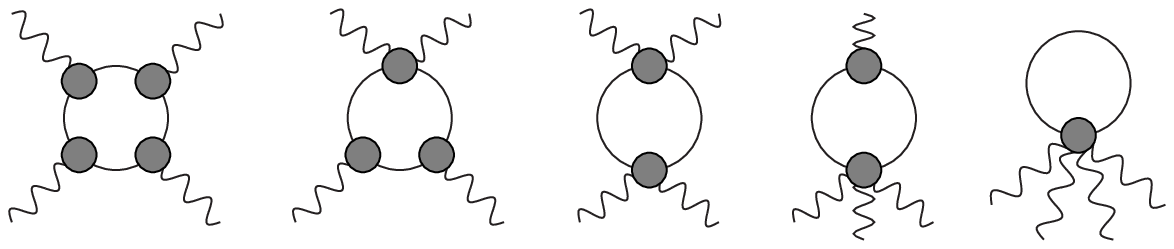}}
\caption{\label{figpiloop}
The pion-loop contributions to the vector four-point function of
Eq.~\ref{defPI}. The modeling is in the expressions for the form-factors
designated by the shaded blobs.
}
\end{figure}

The pion loop contribution is depicted in Fig.~\ref{figpiloop}.
In the models we consider all the diagrams depicted can appear.
The shaded blob indicates the presence of form-factors. In this section we
will only discuss models and not include rescattering and a possible
ambiguity in distinguishing two-pion contributions from scalar-exchanges.
The dispersive method
\cite{Colangelo:2014dfa,Colangelo:2014pva,Colangelo:2015ama} will include
this automatically but at present no full numerical results from this
approach are available.

\subsection{VMD versus HLS}

The simplest model is a point-like pion or scalar QED (sQED). This gives
a contribution of $a_\mu^{\pi loop} \approx -4\cdot 10^{-10}$. However,
at high energies
a pion is clearly not point-like. A first step is to include the pion
form-factor in the vertices with a single photon. Gauge invariance then requires
the presence of more terms with form-factors.
The simplest gauge-invariant addition is to add the pion form-factor
also to both legs of the $\pi\pi\gamma^*\gamma^*$ vertices and neglect vertices
with three or more photons. For the pion form-factor
one can use either the VMD expression or a more model/experimental inspired
version. Using a model for the form-factor, is what was called full VMD
\cite{BPP1,BPP2} and using the experimental data corresponds to what is called
the model-independent or FsQED part of the two-pion contribution in
\cite{Colangelo:2014dfa,Colangelo:2014pva,Colangelo:2015ama}.
The ENJL model used for the form-factor of \cite{BPP1,BPP2} led to
$a_\mu^{\pi loop}\approx-1.9\cdot 10^{-10}$. A form-factor parametrization of the
form $m_V^2/(m_V^2-q^2)$, a VMD parametrization, leads to
$a_\mu^{\pi loop}\approx-1.6\cdot 10^{-10}$ and using the experimental data
FsQED gives $a_\mu^{\pi loop}\approx-1.6\cdot 10^{-10}$ \cite{Procura:2016btl}.

We study which momentum regions contribute most to $a_\mu$ by rewriting
Eq.~(\ref{defintegral}) with integration variables the (Euclidean) off-shellness
of the three photons, $P_1^2,P_2^2,Q^2$. 
In fact to see the regions better we use \cite{BP}
$l_P = (1/2)\ln\left(P^2/\mathrm GeV^2\right)$ for $P=P_1,P_2,Q$.
With these variables we define
\begin{equation}
\label{defaLL}
a_\mu = \int dl_{P_1} dl_{P_2} dl_Q\, { a_\mu^\mathrm{LLQ}}\,.
\end{equation}

\begin{figure}
\centering
\includegraphics[width=14cm]{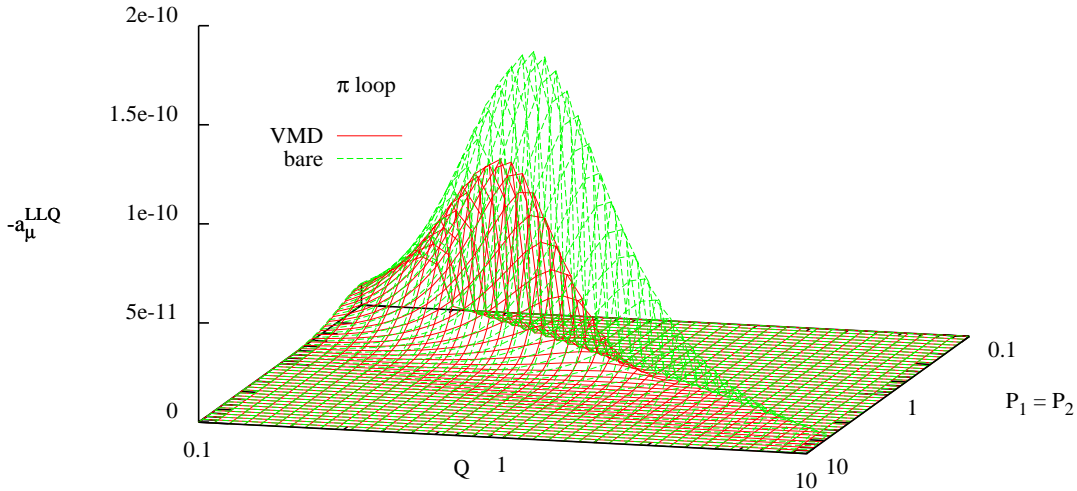}
\caption{The momentum dependence of the pion loop contribution.
Plotted is $a_\mu^{LLQ}$
of (\ref{defaLL})
as a function of $P_1=P_2$ and $Q$. Top surface: sQED, bottom surface:full VMD.}
\label{figpiloopVMD}
\end{figure}

As a first example we show $-a_\mu^{LLQ}$ along the plane with $P_1=P_2$
for the bare pion-loop or sQED and the full VMD in Fig.~\ref{figpiloopVMD}.
The minus sign is included to make the plots easier to see.
The contribution to $a_\mu$ as shown is proportional to the volume under the
surfaces. It is clearly seen how the form-factors have little effect at
low energies but are much more important at high momenta.
We have three variables in principle but we only show plots
with $P_1=P_2$. The reason is that one can see in all our figures that
the results
are concentrated along the line $Q=P_1=P_2$ and fall off fast away from there.
The plots with $P_1\ne P_2$ look similar but are smaller and do
not show anything new qualitatively.

\begin{figure}
\centering
\includegraphics[width=14cm]{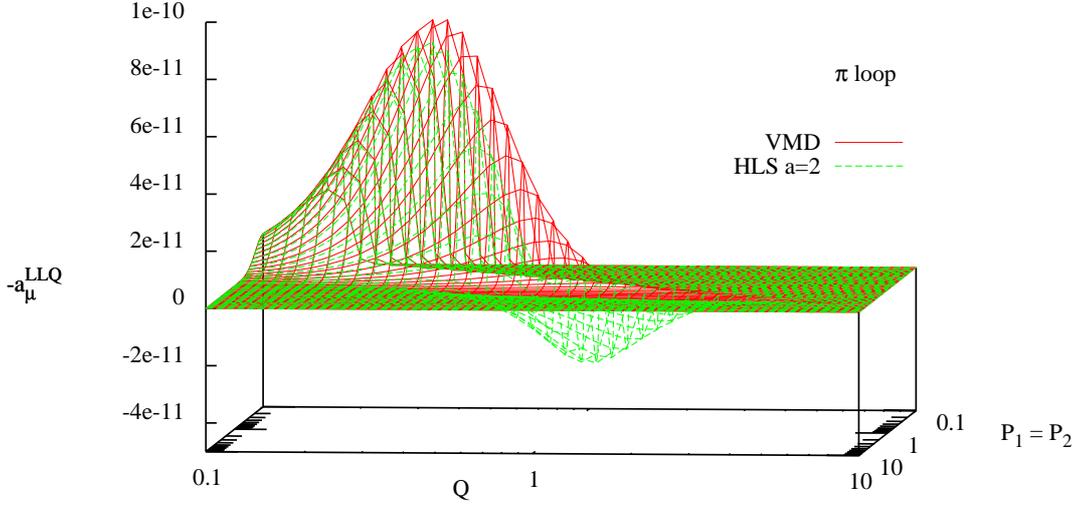}
\caption{$-a_\mu^{LLQ}$
of (\ref{defaLL})
as a function of $P_1=P_2$ and $Q$. Top surface: full VMD, bottom surface: HLS.}
\label{figpiloopVMDHLS}
\end{figure}

The other main evaluation of the pion-loop in
\cite{HKS1,HKS2} (HKS) used a different approach.
It was believed then that the full VMD approach did not respect gauge
invariance. HKS therefore used the hidden local
symmetry model with only vector mesons (HLS) \cite{Bando:1987br} and obtained
$-0.45\times 10^{-10}$. The only difference with full VMD is in
the $\pi\pi\gamma^*\gamma^*$ as discussed in \cite{BPP2}. In \cite{BPP2}
it was shown that the full VMD approach is gauge invariant.
However, the large spread in the results for models that are rather similar
was puzzling, both have a good description of the pion form-factor.
We can make a similar study of the momentum range contributions, shown in
Fig.~\ref{figpiloopVMDHLS}.
It is clearly visible that the two models agree very well for low momenta
but there is a surprisingly large dip of the opposite sign for the HLS
model at higher momenta, above and around 1 GeV. This is the reason for the
large difference in the final number for $a_\mu^{\pi loop}$. A comparison as
a function of the cut-off can be found in \cite{Mehranthesis}.

\subsubsection{Short distance constraint: VMD is better}

In QCD we know that the total hadronic contribution to the muon anomalous
magnetic moment must be finite. This is however not necessarily true
when looking at non-renormalizable models that in addition only describe part
of the total hadronic contribution. For these one has too apply them
intelligently, i.e. only use them in momentum regions where they are valid.

One tool to study possible regions of validity is to check how well the models
do in reproducing short-distance constraints following directly from QCD.
Examples of these are the Weinberg sum rules but there are also some applicable
to more restricted observables. Unfortunately it is known that in general one
cannot satisfy all QCD constraints with a finite number of hadrons included
as discussed in detail in \cite{BGLP}. Still one wants to include as much
as possible of QCD knowledge in the models used.

One constraint on the amplitude for $\gamma^*\gamma^*\to\pi\pi$ can be easily
derived analoguously to the short-distance constraint of \cite{MV} for the
pion exchange contribution.
If we take both photons to be far off-shell and at a similar $Q^2$
then the leading term in the operator product expansion of the two
electromagnetic currents is proportional to the axial current.
However, a matrix element of the axial current with two pions vanishes
so we have the constraint
\begin{equation}
\label{SDconstraint}
\lim_{Q^2\to\infty} A(\gamma^*(q_1=Q+k)\gamma^*(q_2=-Q+k)\to\pi(p_1)\pi(p_2))
\propto \frac{1}{Q^2}
\end{equation}
when all scalar products involving $k,p_1,p_2$ and at most one power of Q
are small compared to $Q^2$.

In scalar QED the amplitude for $\gamma^*\gamma^*\to\pi\pi$ is
\begin{equation}
ie^2\left[2g^{\mu\nu}
  +\frac{(k^\mu+Q^\mu-2p_1^\mu)(k^\nu-Q^\nu-2p_2^\nu)}{(Q+k-p_1)^2-m_\pi^2}
  +\frac{(k^\mu+Q^\mu-2p_2^\mu)(k^\nu-Q^\nu-2p_1^\nu)}{(Q-k+p_1)^2-m_\pi^2}\right]
\label{eq:sQEDamp}
\end{equation}
which to lowest order in $1/Q^2$ is
\begin{equation}
2ie^2\left[g^{\mu\nu} - \frac{Q^\mu Q^\nu}{Q^2}\right].
\label{eq:sQEDampLowest}
\end{equation}
This amplitude does not vanish in the large $Q^2$ limit. sQED does not satsify
the short distance constraint.

In full VMD the $\gamma\pi\pi$ and $\gamma\gamma\pi\pi$ vertices of scalar QED
are multiplied by a factor
\begin{equation}
\frac{m_\rho^2 g^{\mu\nu} - q^\mu q^\nu}{m_\rho^2 - q^2}
\end{equation} 
for each photon line, where $q$ is the momentum of the photon.
The $(Q^2)^0$ term in the $\gamma^*\gamma^*\to\pi\pi$ amplitude is then zero.
The full VMD model does respect the short distance constraint.

In HLS the $\gamma\pi\pi$ vertex of scalar QED is multiplied by
\begin{equation}
g^{\mu\nu} - \frac{a}{2}\frac{q^2 g^{\mu\nu} - q^\mu q^\nu}{q^2-m_\rho^2}
\end{equation}
and the $\gamma\gamma\pi\pi$ vertex is multiplied by
\begin{equation}
g^{\mu\alpha}g^{\nu\beta} - g^{\mu\alpha}\frac{a}{2}\frac{q^2 g^{\nu\beta} - q^\nu q^\beta}{q^2 - m_\rho^2} - g^{\nu\beta}\frac{a}{2}\frac{p^2 g^{\mu\alpha} - p^\mu p^\alpha}{p^2 - m_\rho^2}.
\end{equation}
To lowest order in $1/Q^2$ the amplitude for $\gamma^*\gamma^*\to\pi\pi$ is
\begin{equation}
2ie^2\left[g^{\mu\nu} - \frac{Q^\mu Q^\nu}{Q^2}\right](1-a).
\end{equation}
The HLS model with its usual value of $a=2$ does not satisfy the short distance
constraint.

It was also noticed \cite{BP} in a similar vein that the ENJL
model, that essentially has full VMD, lives up to the Weinberg sum rules
but the HLS does not.

\begin{figure}
\centering
\includegraphics[width=14cm]{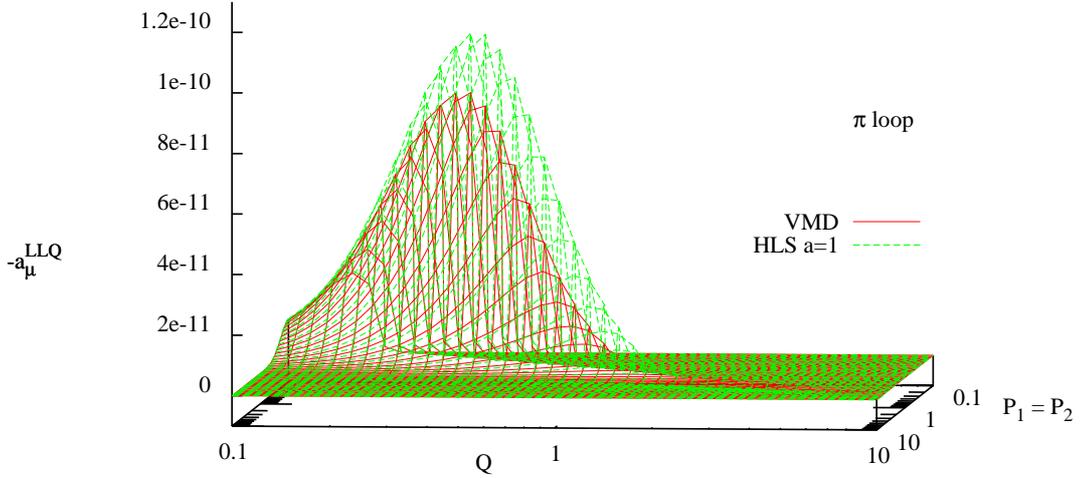}
\caption{The momentum dependence of the pion loop contribution.
$-a_\mu^{LLQ}$
of (\ref{defaLL})
as a function of $P_1=P_2$ and $Q$. Top surface: HLS a=1, bottom surface: full VMD.}
\label{figpiloopVMDHLS2}
\end{figure}

In fact, using the HLS with an
unphysical value of the parameter $a=1$ satisfies the
short-distance constraint (\ref{SDconstraint}) and lives up to the first
Weinberg sum rule. The total result for that model is
$a_\mu^{\pi loop} = -2.1\cdot 10^{-10}$, similar to the ENJL model.
A comparison for different momentum regions between the full VMD model and
a HLS model with $a=1$ is shown in Fig.~\ref{figpiloopVMDHLS2}.
Notice in particular that the part with the opposite sign from
Fig.~\ref{figpiloopVMDHLS} has disappeared.

From this we conclude that a number in the range
$a_\mu^{\pi loop}=-(1.5$-$2.1)\times10^{-10}$
would be more appropriate.

\subsection{Including polarizability at low energies}

\begin{figure}
\centering
\includegraphics[width=14cm]{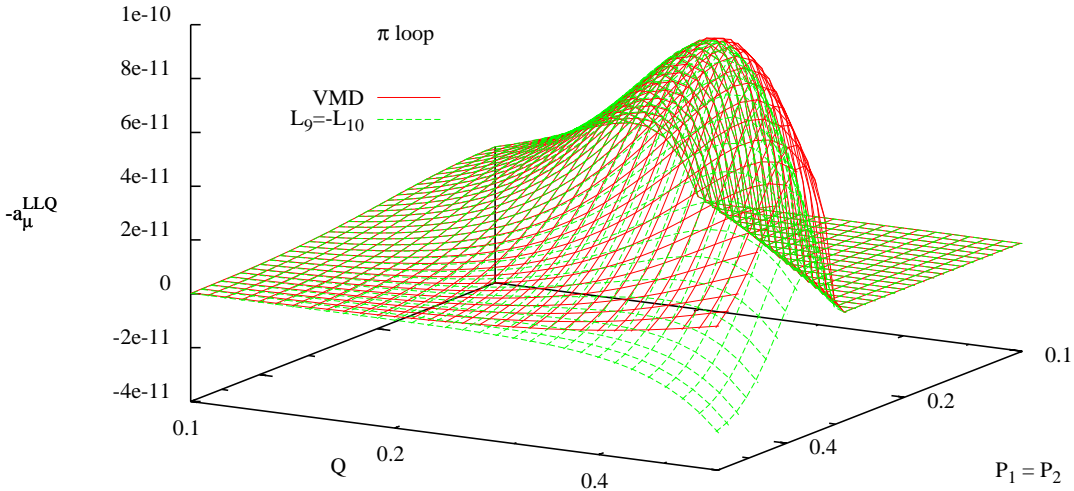}
\caption{$-a_\mu^{LLQ}$
of (\ref{defaLL})
as a function of $P_1=P_2$ and $Q$. Top surface: full VMD, bottom surface: 
ChPT with $L_9=-L_{10}$ so the charge radius is included but no polarizability.}
\label{figpiloopChPT1}
\end{figure}

\begin{figure}
\centering
\includegraphics[width=14cm]{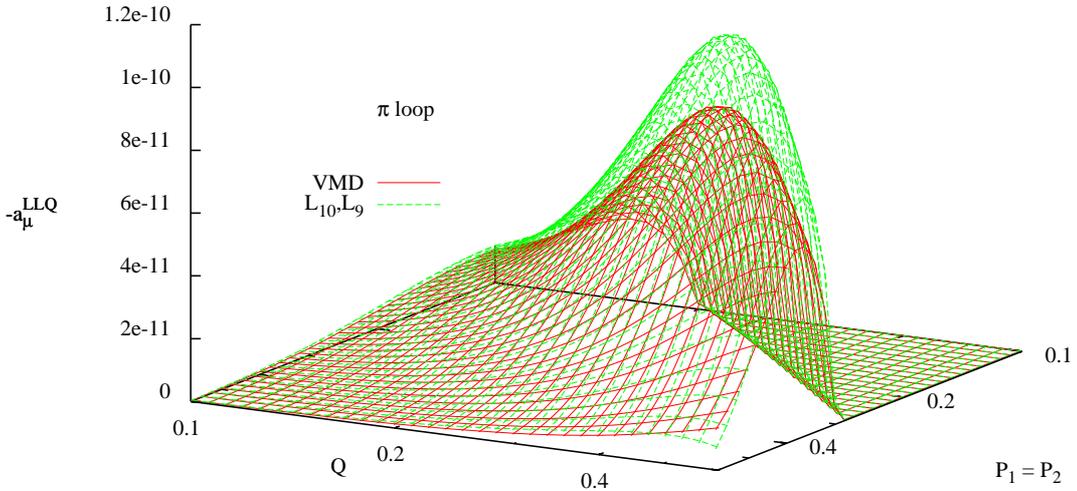}
\caption{$-a_\mu^{LLQ}$
of (\ref{defaLL})
as a function of $P_1=P_2$ and $Q$. Bottom surface: full VMD, top surface: 
ChPT with $L_9\ne-L_{10}$ so the charge radius and the polarizability
are included.}
\label{figpiloopChPT2}
\end{figure}

It was pointed out that the effect of pion polarizability
was neglected in the estimates of the pion-loop in
\cite{HKS1,HKS2,BPP1,BPP2} and a first estimate of this
effect was given using the Euler-Heisenberg four photon effective vertex
produced by pions \cite{Engel:2012xb} within Chiral Perturbation Theory.
This approximation is only valid
below the pion mass. In order to check the size of the pion radius effect
and the polarizability, we have implemented the low energy part
of the four-point function and computed $a_\mu^{LLQ}$ for these cases
in Chiral Perturbation Theory (ChPT).
First results were shown in \cite{talk1,Mehranthesis}. 
The plots shown include the $p^4$ result which is the same as
the bare pion-loop and we include in the vertices the effect of the terms from
the $L_9$ and $L_{10}$ terms in the $p^4$ ChPT Lagrangian. 
The effect of the charge radius is shown in Fig.~\ref{figpiloopChPT1}
compared to the VMD parametrization of it, notice the different momentum
scales compared to the earlier Figs.~\ref{figpiloopVMD}-\ref{figpiloopVMDHLS2}.
The polarizability we have set to zero by setting $L_9+L_{10}=0$.
As expected, the charge radius effect is included in the
VMD result since the latter gives a good description of the pion form-factor.
Including the effect of the polarizability can be done in ChPT by
using experimentally determined values for $L_9$ and $L_{10}$. The latter
can be determined from $\pi^+\to e\nu\gamma$ or the hadronic vector
two-point functions. Both are in good agreement and lead to a prediction
of the pion polarizability confirmed by the Compass experiment
\cite{Adolph:2014kgj}. The effect of including this in ChPT on $a_\mu^{LLQ}$
is shown in Fig.~\ref{figpiloopChPT2}. An increase of 10-15\% over the
VMD estimate can be seen.

ChPT at lowest order, or $p^4$, for $a_\mu$ is just the point-like pion loop
or sQED. At NLO pion exchange with point-like vertices and the pion-loop
calculated at NLO in ChPT are needed. Both give divergent contributions
to $a_\mu$, so pure ChPT is of little use in predicting $a_\mu$. 
If we had tried to extend the plots in Figs.~\ref{figpiloopChPT1} and
\ref{figpiloopChPT2} to higher momenta the bad high energy behaviour would
have been clearly visible. We therefore need to go beyond ChPT. This is done in
the next subsection.

\subsection{Including polarizability at higher energies}

If we want to see
the full effect of the polarizability we need to include a model that can be
extended all the way, or at least to a cut-off of about 1~GeV.
For the approach of \cite{Engel:2012xb} this was done in 
\cite{Engel:2013kda} by including a propagator description of $a_1$
and choosing it such that the full contribution of the pion-loop
to $a_\mu$ is finite. They obtained a range of $-(1.1$-$7.1)\times 10^{-10}$
for the pion-loop contribution. This seems a very broad range when compared
with all earlier estimates.
One reason is that the range of polarizabilities used in \cite{Engel:2013kda}
is simply not compatible with ChPT. The pion polarizability is an observable
where ChPT should work and indeed the convergence is excellent. The ChPT
prediction has also recently been confirmed by experiment \cite{Adolph:2014kgj}.
Our work 
discussed below indicates that $-(2.0\pm0.5)\times10^{-10}$ is a more
appropriate range for the pion-loop contribution.

\begin{figure}
\centering
\includegraphics[width=11cm]{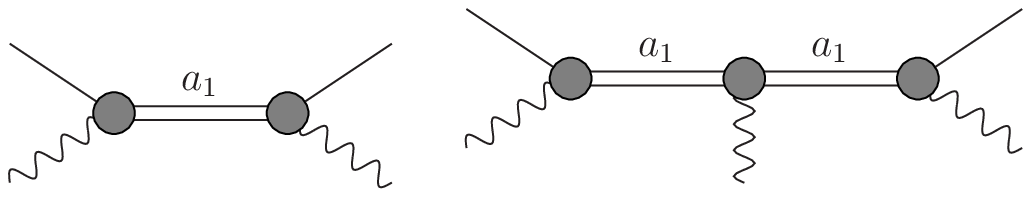}
\caption{Left: the $a_1$-exchange that produces the pion polarizability.
Right: an example of a diagram that is required by gauge invariance.}
\label{figa1pipi}
\end{figure}

The polarizability comes from $L_9+L_{10}$ in ChPT
\cite{Donoghue:1993kw,Bijnens:1987dc}.
Using \cite{Ecker:1988te}, we notice that the polarizability is produced by
$a_1$-exchange depicted in Fig.~\ref{figa1pipi}. This is depicted
in the left diagram of Fig.~\ref{figa1pipi}.
However, once such an exchange is there, diagrams like the right one in
Fig.~\ref{figa1pipi} lead to effective $\pi\pi\gamma\gamma\gamma$ vertices
and are required by electromagnetic gauge invariance. 
This issue can be dealt with in several ways.
Ref.~\cite{Engel:2013kda} introduced modifications of the $a_1$
propagator that introduces one form of the extra vertices.
We deal with them via effective Lagrangians incorporating vector and
axial-vector mesons.

If one studies Fig.~\ref{figa1pipi} one could raise the question ``Is
including a $\pi$-loop but no $a_1$-loop consistent?''
The answer is yes with the following argument. We can first look at a tree
level Lagrangian including pions $\rho$ and $a_1$. We then integrate out the
$\rho$ and $a_1$ and calculate the one-loop pion diagrams
with the resulting all order Lagrangian.
In the diagrams of the original Lagrangian this corresponds to only including
loops with at least one pion propagator present. Numerical results for cases
including full $a_1$ loops are presented as well below.
As a technicality, we use anti-symmetric vector fields for the vector and
axial-vector mesons. This avoids complications due to $\pi$-$a_1$ mixing.
We add vector $V_{\mu\nu}$ and axial-vector $A_{\mu\nu}$ nonet fields.
The kinetic terms are given by \cite{Ecker:1988te}
\begin{equation}
\label{VAkinetic}
-\frac{1}{2}\left\langle\nabla^\lambda V_{\lambda\mu}\nabla_\nu V^{\nu\mu}
-\frac{M_V^2}{2}V_{\mu\nu}V^{\mu\nu}\right\rangle
+ V\leftrightarrow A\,.
\end{equation}
We add first the terms that contribute to the $L_i$ \cite{Ecker:1988te}
\begin{equation}
\label{VALi}
 \frac{F_V}{2\sqrt{2}}\left\langle f_{+\mu\nu}V^{\mu\nu}\right\rangle
+\frac{i G_V}{\sqrt{2}}\left\langle V^{\mu\nu}u_\mu u_\nu\right\rangle
+\frac{F_A}{2\sqrt{2}}\left\langle f_{-\mu\nu}A^{\mu\nu}\right\rangle
\end{equation}
with
$L_9 = \frac{F_V G_V}{2 M_V^2}$, $L_{10}=-\frac{F_V^2}{4M_V^2}+\frac{F_A^2}{4M_A^2}$. The Weinberg sum rules in the chiral limit imply
$F_V^2 = F_A^2 + F^2_\pi$, $F_V^2 M_V^2 = F_A^2 M_A^2$
and requiring VMD behaviour for the pion form-factor $ F_V G_V = F^2_\pi$.
We have used input values for the $L_9$ and $L_{10}$ consistent with this
in the previous subsection.

Calculating the $\gamma^*\gamma^*\to\pi\pi$ amplitude in this framework
using antisymmetric tensor notation to lowest order in $1/Q^2$ gives the
amplitude
\begin{align}
2ie^2&\frac{F_A^2}{Q_1^2 m_a^2 F^2} (  - p_1^\mu Q_1^\nu p_1\cdot Q_1 - p_1^\nu Q_1^\mu p_1\cdot Q_1 + Q_1^\mu Q_1^\nu m_\pi^2 + g^{\mu\nu} (p_1\cdot Q_1)^2 )
\nonumber\\
+ 2ie^2&\frac{F_A^2}{m_a^2 F^2} ( p_1^\mu p_1^\nu - g^{\mu\nu}m_\pi^2)
\nonumber\\
+ 2ie^2&(F_A^2 + F^2 - F_V^2)\left(\frac{g^{\mu\nu}}{F^2} - \frac{Q_1^\mu Q_1^\nu}{Q_1^2 F^2}\right).
\end{align}
The last line vanishes for $F_A^2 + F^2 - F_V^2 = 0$ which is one of
Weinberg's sum rules. However, the first two lines give the additional
requirement $F_A^2 = 0$. In this model it is not possible to incorporate
the $a_1$ meson and satisfy the short distance constraint (\ref{SDconstraint}).

First, we take the model with only $\pi$ and $\rho$, i.e. we only keep the
first two terms of (\ref{VAkinetic}) and (\ref{VALi}).
The one-loop contributions to $\Pi^{\rho\nu\alpha\beta}$
are not finite. They were also not finite for the HLS
model of HKS, but the relevant
$\delta\Pi^{\rho\nu\alpha\beta}/\delta p_{3\lambda}$ was. However, in the present
model, the derivative can be made finite only for
$G_V=F_V/2$. With this value of the parameters the result for $a_\mu$
is identical to that of the HLS model and suffers as a consequences from the
same defects discussed above.

Next we do add the $a_1$ and require $F_A\ne0$. After a lot of work we find
that
$\delta\Pi^{\rho\nu\alpha\beta}/\delta p_{3\lambda}|_{p_3=0}$ is finite only for
$G_V=F_V=0$ and $F_A^2=-2F_\pi^2$ or, if including a full $a_1$-loop
$F_A^2=-F_\pi^2$. These solutions are clearly unphysical.

We then add all $\rho a_1\pi$ vertices given by
\begin{align}
&\lambda_1\left\langle \left[V^{\mu\nu},A_{\mu\nu}\right]\chi_-\right\rangle
+\lambda_2\left\langle \left[V^{\mu\nu},A_{\nu\alpha}\right]{h_\mu}^\nu\right\rangle
\nonumber\\&
+\lambda_3\left\langle i\left[\nabla^\mu V_{\mu\nu},A_{\nu\alpha}\right]u_\alpha\right\rangle
+\lambda_4\left\langle i\left[\nabla_\alpha V_{\mu\nu},A_{\alpha\nu}\right]u^\mu\right\rangle
\nonumber\\&
+\lambda_5\left\langle i\left[\nabla^\alpha V_{\mu\nu},A_{\mu\nu}\right]u_\alpha\right\rangle
+\lambda_6\left\langle i\left[V^{\mu\nu},A_{\mu\nu}\right]{{f_{-}}^\alpha}_\nu\right\rangle
\nonumber\\&
+\lambda_7\left\langle i V_{\mu\nu}A^{\mu\rho}{A^\nu}_\rho\right\rangle\,.
\end{align}
These are not all independent due to the constraints on $V_{\mu\nu}$ and
$A_{\mu\nu}$ \cite{Leupold}, there are three relations.
After a lot of work, we found that no solutions with
$\delta\Pi^{\rho\nu\alpha\beta}/\delta p_{3\lambda}|_{p_3=0}$ exists except
those already obtained without $\Lambda_i$ terms.
The same conclusions holds if we look at the combination that shows up in the
integral over $P_1^2,P_2^2,Q^2$. We thus find no reasonable model
that has a finite prediction for $a_\mu$ for the pion-loop including
$a_1$. In the remainder we therefore stick to $\lambda_i=0$ for the numerical
results.

\begin{figure}
\centering
\includegraphics[width=14cm]{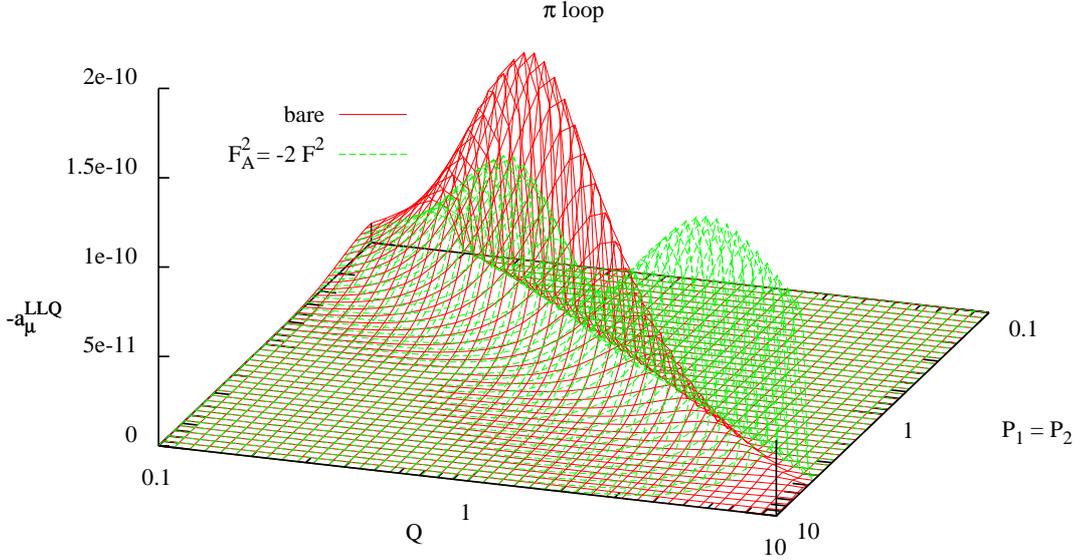}
\caption{$-a_\mu^{LLQ}$ as defined in (\ref{defaLL}) as a function of
$P_1=P_2$ and $Q$ with $a_1$ but no full $a_1$-loop, $F_A^2=-2 F_\pi^2$
and $F_V=G_V=0$.
The bare pion loop is shown for comparison.}
\label{figpiloopfinitenoa1}
\end{figure}

\begin{figure}
\centering
\includegraphics[width=14cm]{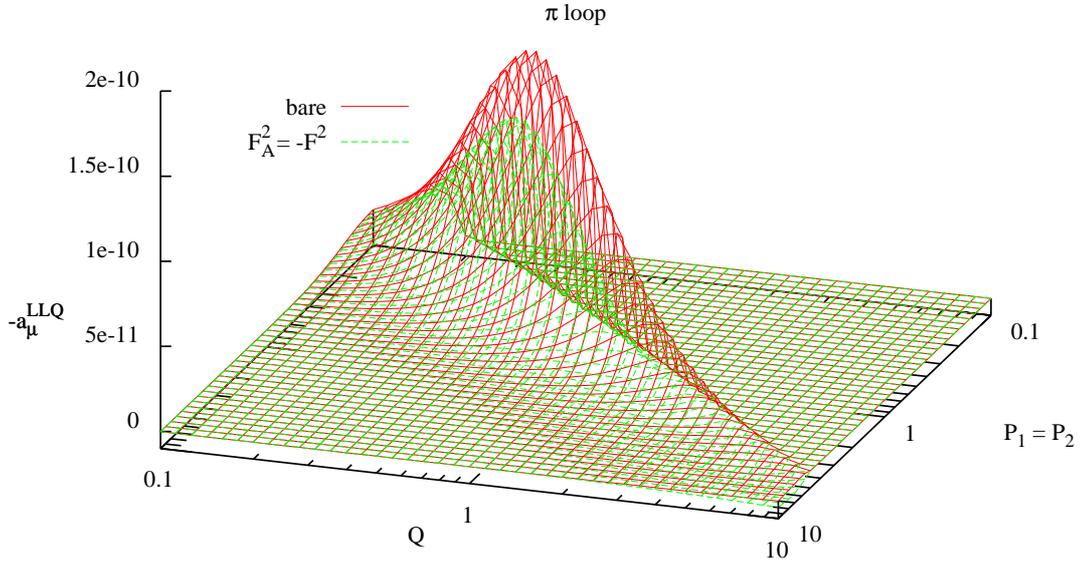}
\caption{$-a_\mu^{LLQ}$ as defined in (\ref{defaLL}) as a function of
$P_1=P_2$ and $Q$ with $a_1$ with a full $a_1$-loop, $F_A^2=- F_\pi^2$
and $F_V=G_V=0$.
The bare pion loop is shown for comparison.}
\label{figpiloopfinitea1}
\end{figure}

Let us first show the result for one of the finite cases, no $a_1$ loop,
$F_V=G_V=0$ and $F_A^2=-2 F_\pi^2$.
The resulting contribution from the different momentum regimes
is shown in Fig.~\ref{figpiloopfinitenoa1}
The high-energy behaviour is by definition finite but there is a large bump
at rather high energies.
The other finite solution, including a full $a_1$-loop
and  $F_A=-F_\pi^2,F_V=G_V=0$ is shown in
Fig.~\ref{figpiloopfinitea1}.
Here the funny bump at high energies has disappeared but the behaviour
is still unphysical. The high-energy behaviour is good by definition since we
enforced a finite $a_\mu$.

\begin{figure}
\centering
\includegraphics[width=14cm]{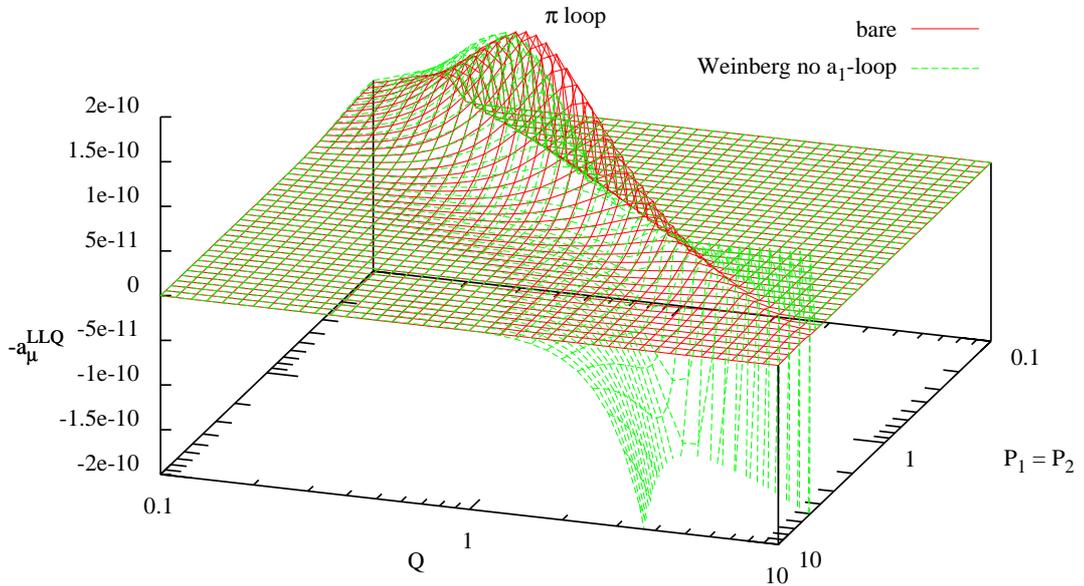}
\caption{$-a_\mu^{LLQ}$ as defined in (\ref{defaLL}) as a function of
$P_1=P_2$ and $Q$ with $a_1$ but no full $a_1$-loop. Parameters determined by the Weinberg sum rules.}
\label{figpiloopweinbergnoa1}
\end{figure}

\begin{figure}
\centering
\includegraphics[width=14cm]{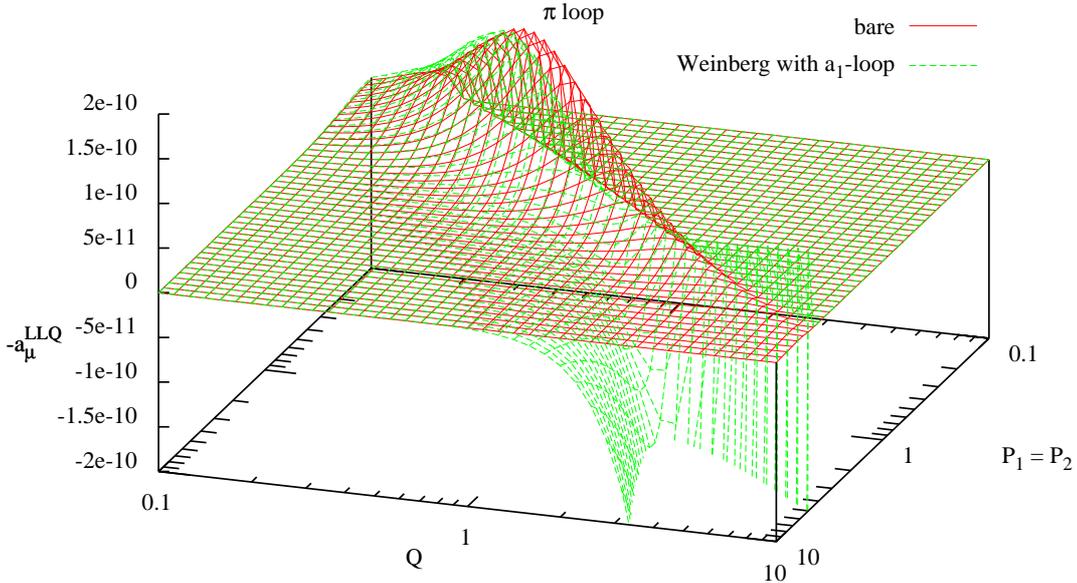}
\caption{$-a_\mu^{LLQ}$ as defined in (\ref{defaLL}) as a function of
$P_1=P_2$ and $Q$ with $a_1$ but no full $a_1$-loop. Parameters determined by the Weinberg sum rules.}
\label{figpiloopweinberga1}
\end{figure}

We can now look at the cases where $a_\mu^{\pi loop}$ was not finite
but that include a good low-energy behaviour. I.e. they have
$F_V^2= F_\pi^2/2$, $F_V G_V=F_\pi^2$, $F_A^2 = F_\pi^2/2$ and $M_A^2 = 2 M_V^2$.
The resulting model then satisfies the Ward identities
and the VMD behaviour of the pion-form factor.
For the case with no $a_1$-loop we obtain $-a_\mu^{LLQ}$ as shown
in Fig.~\ref{figpiloopweinbergnoa1}.
The bad high energy behaviour is clearly visible, but it only starts
above 1~GeV.
The same input parameters but with a full $a_1$-loop
leads to only small changes in the momentum regime considered
as shown in Fig.~\ref{figpiloopweinberga1}
Again the bad high-energy behaviour is clearly visible.

\begin{figure}
\centering
\includegraphics[width=14cm]{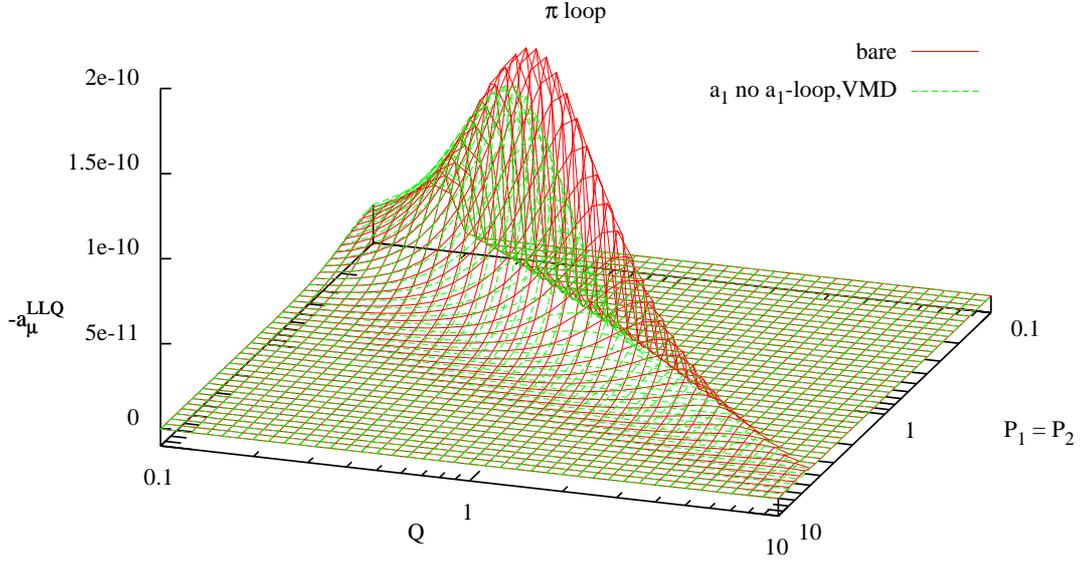}
\caption{$-a_\mu^{LLQ}$ as defined in (\ref{defaLL}) as a function of
$P_1=P_2$ and $Q$ with $a_1$ and $F_A^2=F_\pi^2$ but no full $a_1$-loop.
A VMD form-factor is added in all photon legs.}
\label{figpiloopnoa1VMD}
\end{figure}

\begin{figure}
\centering
\includegraphics[width=14cm]{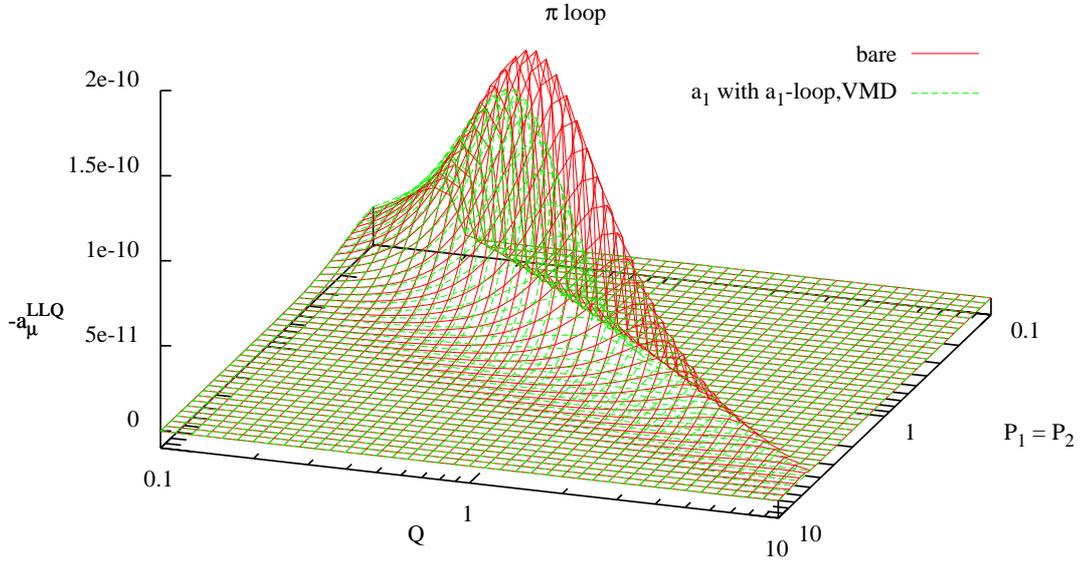}
\caption{$-a_\mu^{LLQ}$ as defined in (\ref{defaLL}) as a function of
$P_1=P_2$ and $Q$ with $a_1$ and $F_A^2=F_\pi^2$ with a full $a_1$-loop.
A VMD form-factor is added in all photon legs.}
\label{figpiloopa1VMD}
\end{figure}

As a last model, we take the case with $F_A^2= + F_\pi^2$ and add
VMD propagators also in the photons coming from vertices involving $a_1$.
This makes the model satisfy the short-distance
constraint (\ref{SDconstraint}).
The contributions to $a_\mu^{\pi loop}$ are shown in
Fig.~\ref{figpiloopnoa1VMD}.
The same model but now with the full $a_1$-loop is shown in
Fig.~\ref{figpiloopa1VMD}.
Both cases are very similar and here is a good high energy behaviour due to the
VMD propagators added. This model cannot be reproduced by the
Lagrangians shown above, we need higher order terms to do so.
However, the arguments of \cite{BPP2} showing that
the full VMD model was gauge invariant also apply to this model.

\begin{figure}[tb]
\centering
\includegraphics[width=14cm]{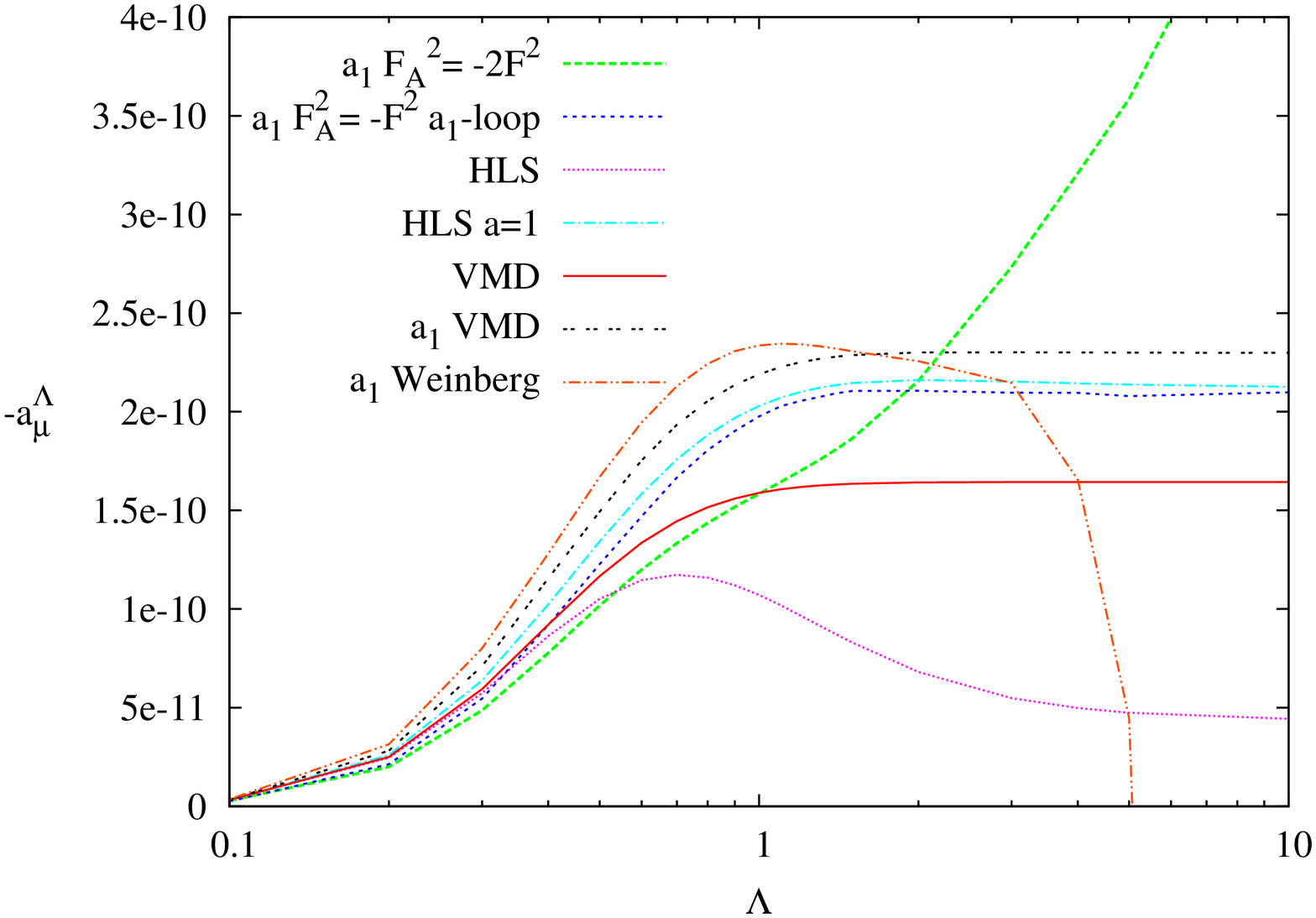}
\caption{$-a_\mu$ using a variety of models for the pion loop as a function of $\Lambda$, the cut-off on the photon momenta. Units for $\Lambda$ are GeV.}
\label{figpiloopall}
\end{figure}

Now how does the full contribution to $a_\mu^{\pi loop}$ of these various
models look like. The integrated contribution up to a maximum $\Lambda$
for the size of $P_1,P_2$ and $Q$ is shown in
Fig.~\ref{figpiloopall}.
The models with good high energy behaviour are the ones with a horizontal
behaviour towards the right. We see that the HLS is quite similar to the
others below about 0.5 GeV but then drops due to the part with the sign
as shown in Fig.~\ref{figpiloopVMDHLS}.
All physically acceptable models that show a reasonable enhancement over the
full VMD result. In fact,
all models except HLS end up with a value of $a_\mu=-(2.0\pm0.5)\times10^{-10}$
when integrated up-to a cut-off of order 1-2~GeV.
We conclude that that is a reasonable estimate for the pion-loop
contribution.

We have not redone the calculation with the model of \cite{Engel:2013kda},
however their large spread of numbers comes from considering a very broad
range of pion polarizabilities and we suspect that the result might contain
a large contribution from high energies similarly to the model shown in
Fig.~\ref{figpiloopfinitenoa1}.  We therefore feel that their broad range should
be discarded.

\section{Summary and conclusions}
\label{Conclusions}

In this paper we have two main results and two smaller ones.
The first main result is that we expect large and opposite sign contribution
from the disconnected versus the connected parts in lattice calculations
of the HLbL contribution to the muon anomalous magnetic moment.

The second main result is that the estimate of the pion-loop is 
\begin{equation}
\label{resultpiloop}
a_\mu^{\pi loop}= -(2.0\pm0.5)\cdot 10^{-10}\,.
\end{equation}
This contains the effects of the pion polarizability as well as estimates
of other $a_1$ effects. The main constraints are that a realistic limit to
low-energy ChPT seems to constrain the models enough to provide the result
and range given in (\ref{resultpiloop}).
We have given a number of arguments why the HLS number
of \cite{HKS1,HKS2} should be considered obsolete.
In this context we have also derived a short distance constraint on the
underlying $\pi\pi\gamma^*\gamma^*$ amplitude.

As a minor result we have given the extension of the Gegenbauer polynomial
method of \cite{KN,JNreview} to the most general hadronic vector four-point
function.
 
\section*{Acknowledgements}

We thank Mehran Zahiri Abyaneh who was involved in the early stages of
this work. This work is supported in part by the Swedish Research Council grants
contract numbers 621-2013-4287 and 2015-04089 and by
the European Research Council (ERC) under the European Union's Horizon 2020
research and innovation programme (grant agreement No 668679).

\end{document}